# Photonic Floquet time crystals

Bing Wang[†,1], Jiaqi Quan[†,2], Jianfei Han[3], Xiaopeng Shen[*,3], Hongwei Wu[*,2], and Yiming Pan[*,4]

1. National Laboratory of Solid State Microstructures and School of Physics, Nanjing University, Nanjing 210093, CHINA.

2. School of Mechanics and Photoelectric Physics, Anhui University of Science and Technology, Huainan 232001, CHINA.

3. School of Materials Science and Physics, China University of Mining and Technology, Xuzhou 221116, CHINA.

4. Physics Department and Solid State Institute, Technion, Haifa 32000, ISRAEL.

† These authors contributed equally to this work.

*Corresponding author. Email: yiming.pan@campus.technion.ac.il; hwwu@aust.edu.cn; xpshen@cumt.edu.cn

**Abstract:**

**The public and scientists constantly have different perspectives. While on a time crystal, they stand in line and ask: What is a time crystal? Show me a material that is spontaneously crystalline in time? This study synthesizes a photonic material of Floquet time crystals and experimentally observes its indicative period-$2T$ beating. We explicitly reconstruct a discrete time-crystalline ground state and reveal using an appropriately-designed photonic Floquet simulator the rigid period-doubling as a signature of the spontaneous breakage of the discrete time-translational symmetry. Unlike the result of the exquisite many-body interaction, the photonic time crystal is derived from a single-particle topological phase that can be extensively accessed by many pertinent nonequilibrium and periodically-driven platforms. Our observation will drive theoretical and technological interests toward condensed matter physics and topological photonics, and demystify time crystals for the non-scientific public.**

**One Sentence Summary:** We reported a first photonic material of topological Floquet time crystals in optics and observed its topologically protected period-doubling.

**Main Text:**

A time crystal is an exotic nonequilibrium state of matter that repeats itself in space and time and spontaneously breaks the continuous or discrete time-translational symmetry. It mimics an ordinary crystal's ability to spontaneously break its spatial translation symmetry in space dimension when cooling down. In 2012, Wilczek first proposed the concept of time crystals in both the classical and quantum variants (*1-3*); however, in the subsequent years, the original time-crystalline model was proved to be invalid in static equilibrium and later ruled out by a no-go theorem (*4, 5*). Moreover, breakage of the discrete time-translational symmetry in periodically driven systems has not been ruled out (*6-9*), and it is a promising research direction (*10-15*). To date, most materialized time crystals follow this direction and are rapidly explored in a number of quantum simulation platforms such as trapped ions (*16*), diamond nitrogen-vacancy centers (*17*), superfluid quantum gases (*18, 19*), and nuclear magnetic resonances (*20, 21*). However, the ground state (a state of minimum energy) of a genuine time-crystalline phase is elusive because it is intrinsically non-conservative out-of-equilibrium after its time-translational symmetry is spontaneously broken.

Usually, the existence of a spontaneous time-crystalline phase can be assessed by probing a stable subharmonic response in experiments, such as period doubling (with twice the period of the underlying drive). However, time crystals are not the only materials or systems that can give rise to subharmonic period-$2T$ oscillation (*16, 22*). Indeed, even before the inception of the discrete time-crystalline phase, period-$2T$ oscillations have been widely studied in a variety of classical and quantum systems, such as period-doubling bifurcation from a logistic map (*23*), subharmonic response in chaotic (*24-26*) or dissipative (*10, 27, 28*) systems, and coupled nonlinear parametric oscillators (e.g., Van der Pol oscillator) (*22, 26*). Moreover, several fascinating candidates for time-crystalline systems have been proposed recently (*3, 18, 29-33*).

In other words, the crucial subharmonic hard evidence required to experimentally substantiate a time-crystalline phase is disputed.

To address the debate, the following question must be posed: Does the time-crystalline phase of matter possess a single-particle excitation, many-body collective phenomenon, or chaos-induced dynamic behavior? No singular answer to this question has been proposed. To explore the actual phenomenon of time crystals, more experiments are needed. Hence, this paper explores the construction of a single-particle picture of time crystals that enables a rigid period-$2T$ oscillation (*34, 35*). Inspired by the discoveries of topological insulators (*36, 37*) and topological photonics (*38*), we attempt to experimentally realize a robust photonic Floquet time crystal that can facilitate the development of periodically driven time-crystalline topological phases. Our observation indicates the universal presence of Floquet time-crystalline states in topological materials and systems (*14, 15, 34, 35*).

**Results -** Let us first formulate an analytical expression to elucidate the ground state of a Floquet or discrete time crystal (FTC, or DTC). To illustrate the spin chain, as schematically shown in Fig. 1A, a many-body-localization-enabled time-crystalline state (i.e., $\pi$-spin glass) was intensively studied (*7-9, 39*), given by $|\{\pm\}\rangle = (|\uparrow\downarrow\uparrow\uparrow\uparrow\uparrow\downarrow\uparrow \cdots \rangle \pm |\downarrow\uparrow\downarrow\downarrow\downarrow\downarrow\uparrow\downarrow \cdots \rangle)/\sqrt{2}$ (also called Schrödinger's cat states). Notably, the $\pi$-spin glass states $|\{\pm\}\rangle$ are the eigenstates of the Floquet evolution operator $U_F|\{\pm\}\rangle = e^{-i\epsilon_{\pm}T}|\{\pm\}\rangle$; consequently, period-doubling was expected due to energy splitting, i.e., $|\epsilon_+ - \epsilon_-| = \pi/T$, during a driving period $T$. This fully entangled many-body ground state is dubbed "$\pi$SG/DTC" ($\pi$-spin glass/discrete time crystal) (*9*). By emulating the $\pi$SG/DTC Floquet states, a many-body Floquet product state can be conjectured, suggested as

$$|\Psi_{FTC}\rangle = |edge\ state\rangle \otimes |DW\rangle \otimes |DW\rangle \otimes \cdots |DW\rangle \otimes |edge\ state\rangle, \quad (1)$$

where the edge state $\left(\sim \frac{|\uparrow\rangle \pm |\downarrow\rangle}{\sqrt{2}}\right)$ is the disorder-protected quantum order at the ends, and the domain walls (DWs) $\left(\sim \frac{|\uparrow\downarrow\rangle \pm |\downarrow\uparrow\rangle}{\sqrt{2}}\right)$ are the protected local integral of motions (LIOMs) randomly distributed on the chain (*39, 40*). Thus, the constructed many-body time-crystalline state (Eq. 1) consists of the direct product of these local single-particle topological edge states and domain walls, as shown in Fig. 1B.

Consider a topological Floquet phase holding both zero- and $\pi$-quasienergy modes at the ends or in the kink structures (*34, 41*). Correspondingly, the edge-state and DW excitations can be described as the superposition states of two topological modes, i.e.,

$$|edge\ state\rangle = \frac{|0\rangle \pm |\pi\rangle}{\sqrt{2}}, |DW\rangle = \frac{|0DW\rangle \pm |\pi DW\rangle}{\sqrt{2}}. \tag{2}$$

As shown in Fig. 1B, a period-$2T$ beating occurs due to the superposition of zero ($|0\rangle, |0DW\rangle$) and $\pi$ quasienergy eigenstates ($|\pi\rangle, |\pi DW\rangle$). The dynamic intensity of a local superposition Floquet state is prototypically given by $I_\pm(x,t) = |\psi_0(x,t) \pm \psi_\pi(x,t)|^2 = |\psi_0(x,t)|^2 + |\psi_\pi(x,t)|^2 \pm 2\Re\{\psi_0^*(x,t)\psi_\pi(x,t)\}$. According to the Floquet–Bloch theorem (*42, 43*), the two Floquet eigenstates are given by $\psi_{0,\pi}(x,t) = u_{0,\pi}(x,t)e^{-i\epsilon_{0,\pi}t}$, where $u_{0,\pi}(x,t) = u_{0,\pi}(x,t+T)$ defines the micromotion within one cycle and $\epsilon_{0,\pi}$ are the corresponding quasienergies. Therefore, the intensity of superposition is obtained as follows:

$$I_\pm(x,t) = |u_0(x,t)|^2 + |u_\pi(x,t)|^2 \pm 2\Re\{u_0^*(x,t)u_\pi(x,t)\}\cos(|\epsilon_0 - \epsilon_\pi|t). \tag{3}$$

Period-doubling occurs if the quasienergy difference $|\epsilon_0 - \epsilon_\pi| = \pi/T$ is locked. We can easily corroborate the relation $I_\pm(x, t+2T) = I_\pm(x,t)$, so that the intensity evolves with a double period of the drive, as shown in Fig. 1C. Namely, the $2T$-periodic subharmonic response stems from the interference between the symmetry-protected Floquet eigenmodes.

Two facets are worth mentioning here. First, it is not difficult to pick two quasienergy states with the splitting $\pi/T$ in the Floquet–Brillouin zone $\left(-\frac{\pi}{T}, \frac{\pi}{T}\right)$ and realize the period-$2T$ beating. However, the problem is that this type of subharmonic response is fragile as it can be destroyed by an infinitely small perturbation. The vulnerability of this period-doubling reflects that the splitting can be quickly affected by other Floquet eigenstates, and period-$2T$ interference (the third term in Eq. 3) is perturbed. Given this issue of stability, only the nontrivially gapped Floquet modes are immune against scattering from imperfections and disorders (*34*).

Second, the relevant subharmonic interference must be local, i.e., $2\Re\{u_0^*(x,t)u_\pi(x,t)\} \neq 0$, and consequently the 0 or $\pi$ edge states and DWs are located at the same ends or kinks. The micromotions of the different spatially located 0 and $\pi$ modes are too far apart to overlap, yielding no interference. Thus, only the local superposition between 0 and $\pi$ modes predominantly contributes to the emergent period-$2T$ oscillation.

**Modeling and setup** - The photonic FTC was modeled after the Su–Schrieffer–Heeger (SSH) model for polyacetylene (1979) (*44*), which has been widely investigated on many photonic simulation platforms (e.g., (*38*)). The edge states and DWs can be modeled in a driven SSH model for electrons (Fig. 1D). For easy implementation in photonic systems, we chose the periodically driven SSH chain. The Hamiltonian of this biatomic model is given by $H(t) = \sum_{i=1}^{N-1}[\kappa_0 + (-1)^i(\delta\kappa_0 + \delta\kappa(t))]c_i^\dagger c_{i+1} + h.c.$, where $c_i^\dagger(c_i)$ are the creation (annihilation) operators of the light field amplitude on the $i^{th}$ waveguide.

The time-periodic coupling term between two nearest-neighboring waveguides (or sites) was dimerized, $\kappa_{i,i+1}(t) = \kappa_0 + (-1)^i(\delta\kappa_0 + \delta\kappa(t))$, where $\kappa_0$ is the constant coupling strength, and $\delta\kappa_0$ and $\delta\kappa(t)$ are the time-independent staggered coupling strengths due to global dimerization and time-periodic dimerization, respectively. $\delta\kappa(t) = \delta\kappa_1\cos(\omega t + \theta)$, where $\delta\kappa_1$ is the strength of the coupling, $\omega = 2\pi/T$ is the Floquet driven frequency, and $\theta$ is the initial phase (Floquet gauge) of the drive. Equivalently, the topological phases in the driven SSH model can be mapped onto the transverse field Ising model (7) or the Kitaev model for a p-wave superconductor (14, 35).

In this experiment, the periodic coupling $\kappa_{i,i+1}(t)$ was appropriately designed and fully controlled by the spatial spacing ($G$) between two neighboring curved waveguides. As shown in Fig. 1D, a photonic simulator of driven SSH chain was designed by mapping the evolution time $t$ of an electron in the direction of light propagation $z$, and correspondingly mapping the Floquet cycle $T$ onto the curving period $\Lambda$. Thus, the coupling profile was given by $\kappa_{i,i+1} = \kappa_{i,i+1}(G, \Lambda)$. For demonstration, we defined an effective coupling length $l_c = \pi/2\kappa_0$ of the simulator to compare it with the period $\Lambda$. The dimerization conditions required for further fabrication were $\delta\kappa_0 \ll \kappa_0$ and $\delta\kappa_1 \ll \kappa_0$. Typically, the coupling length $l_c$ is in the range of 20–100 mm, and the curving period is fixed to $\Lambda = 100$ mm. The coupling profiles extracted from the simulations can be found in the Supplementary Material file.

**Quasienergy gap opening -** Figure 2 demonstrates the quasienergy band for the emergent topological phase coexistence of both zero and $\pi$ Floquet modes. The quasienergy spectrum was precisely calculated using the eigenvalue problem analyses of the Floquet Hamiltonian $H_F = \frac{i}{\Lambda} exp\left(-i \int_0^\Lambda H(z)\,dz\right)$, in which the Hamiltonian is time-periodic $H(z) = H(z + \Lambda)$. Fig. 2A presents the desired band as a function of the Floquet cycle ($\Lambda$) with respect to the effective coupling length ($l_c$). Given the two dimerizations ($\delta\kappa_0 \neq 0, \delta\kappa_1 \neq 0$), the two Floquet modes coexist at the periodically curving condition $\Lambda/l_c \in (1, 2)$, which is associated with two quasienergy gap invariants (41) (see SM file).

Correspondingly, in the region of coexistence, eigenstates of the 0 and $\pi$ modes are illustrated in Fig. 2B. We plotted their evolutionary patterns over four cycles ($4T$), showing that both Floquet modes are periodic in $T$. The intensity of the 0 mode is mainly localized on the first waveguide of the boundary of the array. The $\pi$-mode periodically propagates along with the first two waveguides. Notably, the sole eigenstate excitation of either 0 mode or $\pi$ mode cannot produce the period-doubling oscillation.

Figures 2C and 2D demonstrate the dependence of zero- and $\pi$-gap opening on the staggered strengths $\delta\kappa_0$ and $\delta\kappa_1$, respectively. In a fixed cycle in the Floquet coexistence region, the zero- and $\pi$-gaps closed at $\delta\kappa_0 = 0$ and $\delta\kappa_1 = 0$, separately, as depicted in the insets, and then opened linearly as the dimerization strengths increased. The difference between the two modes is that, while the 0 mode would disappear at $\delta\kappa_0 < 0$ (corresponding to a trivial phase), the $\pi$ mode still exists at $\delta\kappa_1 < 0$. The reason is that for the negative periodic staggered coupling strength, $-\delta\kappa_1 \cos(\omega t + \theta) = \delta\kappa_1 \cos(\omega t + \theta + \pi)$ corresponds to the $\pi$-phase shift for a Floquet gauge choice. That is, the $\pi$ modes emerging with the negative dimerization coupling $(-\delta\kappa_1)$ and gauge $(\theta)$ are equivalent to the modes with positive dimerization coupling $(+\delta\kappa_1)$ and gauge $(\theta + \pi)$. Therefore, the advantage of this array design is that the global dimerization and time-period dimerization solely control the opening of the 0- and $\pi$-gaps, respectively. This dependence made it convenient to control and demonstrate our theoretical expectations using the photonic simulation and measurement.

**Period-2$T$ beating in photonic Floquet simulator -** The photonic simulator was made up of coupled ultrathin corrugated copper strips that support spoof surface plasmon polaritons (SPPs) propagating at microwave frequencies as the highly confined guided wave on a plasmonic waveguide. These ultrathin microstrip lines were first proposed by (*45*) and were recently implemented as a simulation platform to observe anomalous topological modes (*46*). These ultrathin waveguides were deposited on a flexible dielectric substrate (F4BK) that can be bent, folded, coiled, and twisted to guide the spoof SPPs (*45*). Prior to the array fabrication, we used a finite element method in a commercial software (COMSOL Multiphysics) to numerically simulate the near-field distribution of the TM-polarized wave (perpendicular to the ultrathin metallic waveguide interface) propagating along the *z*-direction on the proposed array. For details regarding the fabrication, measurement, and simulation, see SM file.

Figure 3 compares the theoretical expectation, numerical simulation, and experimental observation of the time-crystalline stroboscopic evolution of topological superposition. As observed from the experiment in Fig. 3C, the period-$2T$ oscillation in the curved array sample was probed, and the results perfectly agreed with the prediction (Fig. 3A) and simulation (Fig. 3B). The array consists of ten curved waveguides ($N = 10$) with a period length (Floquet cycle) $\Lambda = 100$ mm and a total length $L = 400$ mm. Its structural fabrication mimics the periodically driven SSH model by considering both global and time-periodic dimerizations. The experimental near-field intensity pattern was scanned by a metallic tip to detect the electric field ($E_z$) over the simulator surface, which was connected to a network analyzer to collect data and perform near-field distribution imaging (see SM file).

With the coupling profiles corresponding to the structural parameters, we calculated the dynamic evolution from the first waveguide of the array. Fig. 3A illustrates the intensity distribution within four Floquet cycles ($4T$), demonstrating the period-$2T$ oscillation twice along the array's boundaries. As observed in the quasienergy band, the input field was mainly projected on to the chiral-symmetry-protected edge modes ($|\psi_{in}\rangle = \alpha|0\rangle + \beta|\pi\rangle$). After $N$ Floquet cycles, the stroboscopic dynamics of the superposition state were given by $U_F(NT)|\psi_{in}\rangle = \alpha|0\rangle + (-1)^N \beta|\pi\rangle$. Thus, the resulting period-doubling appears because the period number ($N$) must be even for the renewal of its initial state ($|\psi_{in}\rangle$).

**Single-particle versus many-body time crystals** - The time-crystalline subharmonic response appears on the boundaries of a chain and can emerge in the bulk as DWs (as the interfacial modes between Floquet phases). To mimic the many-body Floquet time-crystalline phase (Eq. 2), we designed a Floquet simulator with DWs and edge states,

$$|edge\ state\rangle \otimes |DW\rangle \otimes |edge\ state\rangle.$$

As depicted in Fig. 1D, the two Floquet edge states lie at the opposite ends, and a single DW is positioned in the middle of the array. The existence of DWs depends only on the global topological difference of both sides (*34*), while it is free from the local coupling profiles. The principle of bulk-edge correspondence supports DW excitations (*36*). To retain the structural symmetry of the photonic simulator, the middle waveguide was set as straight (Fig. 4A). Nevertheless, the local central couplings to the laterally curved waveguides were still periodically modulated. We conclude that both the end states and DWs permit the coexistence of the two anomalous Floquet modes in this design.

Figures 4B-4D show the simulated field distributions of the many-body time-crystalline states with three inputs: central DW, two end states, and their combination. The input frequency was 17.0 GHz and divided into four cycles, while the array length was the same 400 mm. The waveguide number was increased up to $N = 15$ with a central straight waveguide on the $10^{th}$ site to hold the DW. In this simulation, the field distribution propagated along the waveguides oscillating with a period twice that of the curving. The local period-doubling behavior was rigid, owing to the spatial separation of the topological excitations, so that it does not suffer finite-size hybridization. By adding more randomly distributed yet adequately separated DWs in the Floquet simulator, to reasonably extend, we are able to observe a many-body time-crystalline oscillation (Eq. 1).

To explain the many-body product state dephasing into a fully entangled state, we should consider the kink interactions between the DWs (*39, 40, 47*). This interaction represents renormalized, exponentially decaying interactions between the local integrals of motions in the context of many-body localization (*39*). Then, we can expect a modest kink–kink scattering to fade the product state (Eq. 1) into an entangled many-body state, rather than violating the local DW excitations. As an example, a resulting fully entangled state is written as

$$|\Psi'_{FTC}\rangle = \frac{|\cdots \Uparrow\Downarrow\Uparrow\Uparrow\Uparrow\Downarrow \cdots\rangle \pm |\cdots \Downarrow\Uparrow\Downarrow\Downarrow\Downarrow\Uparrow \cdots\rangle}{\sqrt{2}}, \quad (4)$$

where these local 'spins' are effectively defined as $|\Uparrow\rangle = (|0DW\rangle + |\pi DW\rangle)/\sqrt{2}$, $|\Downarrow\rangle = (|0DW\rangle - |\pi DW\rangle)/\sqrt{2}$ (Eq. 2). This many-body state is an example where all local topological excitations are fully entangled. It can be directly compared with the many-body-localization-enabled $\pi$SG/DTC state (*7-9*). Here, we suppressed the notation of the edge states at the ends in Eq. 4, which resembles the disorder-induced quantum order in the MBL phase (*39*). In view of the spatial confinement characteristics of the entangled Floquet modes, only the local 0 and $\pi$-mode superposition in the effective spins permits the subharmonic interference (Eq. 3). Therefore, the entangled state $|\Psi'_{FTC}\rangle$ also experiences a signature period-$2T$ oscillation. For photonic realization, mimicking the many-body entanglement requires the waveguides to be optically nonlinear (*48*). This exceeds the capability of the proposed Floquet simulator because the nonlinearity of the spoof plasmonic waveguides is negligible (*45*). With this limitation in mind, we realized that it would be difficult to demonstrate the quantum dephasing and entanglement using the proposed simulators. Therefore, we leave this challenge to future research on other quantum many-body platforms (*16-21*).

**Topological Floquet phase transition -** A close look at the quasienergy band outside the coexistence region (Fig. 2A) reveals that only the zero modes survive in the high-frequency-driven region $\Lambda/l_c \in (0, 1)$, and only the $\pi$-mode appears in the intermediate region $\Lambda/l_c \in (2, 3)$. By decreasing the curving period, sequent harmonic (period-$T$), subharmonic (period-$2T$), and static responses can be expected on the boundaries because the corresponding Floquet systems undergo a phase transition from the $\pi$ mode, topological phase coexistence, and zero modes (*34*). For sample fabrication, a simpler way to detect this transition is by altering the ratio between the global and periodic dimerizations.

Figure 5 demonstrates the experimental near-field observation of the topological transition from a harmonic ($T$) and subharmonic ($2T$) to a static response. Fig. 5A illustrates the $\pi$-mode excitation, compared with the case $\delta\kappa_0 = 0$ in Fig. 2C. Then, we increased the global staggered distance between two neighboring waveguides. Zero-modes emerged owing to the nontrivial zero-gap opening via increasing global dimerization. Likewise, Fig. 5B illustrates the subharmonic periodic-$2T$ evolution owing to the coexistence of the two modes. Finally, at the almost fully dimerized limit (corresponding to no coupling between the first two waveguides), only the first waveguide can propagate through the input field, indicating the existence of an extremely isolated zero mode, but the $\pi$ mode is suppressed. As shown in Fig. 5C, most of the input fields remain on the first waveguide, demonstrating the zero-mode isolation. However, few input fields diffuse into the array owing to the effective surplus residual coupling between waveguides. In addition, we thoroughly investigated the dependence of Floquet cycles and input frequencies in the Supplementary Material file.

**Conclusion -** We designed a photonic-material alternative of Floquet time crystal and observed its prototypical period-doubling behavior in our Floquet simulators. To avoid many-body interaction in the photonic simulation, we reconstructed a topologically protected time-crystalline state composed of Floquet topological edge states and domain walls. In a first, both single-particle and many-body pictures of discrete time-crystalline phases were demonstrated.

We believe that the noninteracting topological Floquet time crystals can be easily extended for implementation on many classical and quantum simulation platforms. Also, we hope that the photonic Floquet time crystal can shed light on exotic time-crystalline phase transitions and spur the further development of the out-of-equilibrium state of matter in photonics and condensed matter.

**Acknowledgements**

**Funding:** This work was supported by the German–Israeli DIP Program, by an Advanced Grant from the European Research Council and by the Israel Science Foundation, and by National Natural Science Foundation of China (NSFC) (Grants No. No. 11904008), National Natural Science Foundation of China (61372048), and also the Six Talent Peaks Project in Jiangsu Province of China (XYDXX-072). **Author contributions:** Y. P. and B. W. proposed the concept and modeling. J. Q. and H. W. contributed to the numerical simulation, design, and setting of the experiment. J. H. and X. S. performed the sample fabrication and near-field measurement and analyzed them all together with others. YP drafted the manuscript with all authors' significant contributions to this work. **Competing interests:** The authors declare no competing interests. **Data and materials availability:** All data are available in the main text or the supplementary materials.

**Supplementary Materials**

The supplementary materials are available at [website].

Materials and Methods

Supplementary text.

Figs. S1-S6.

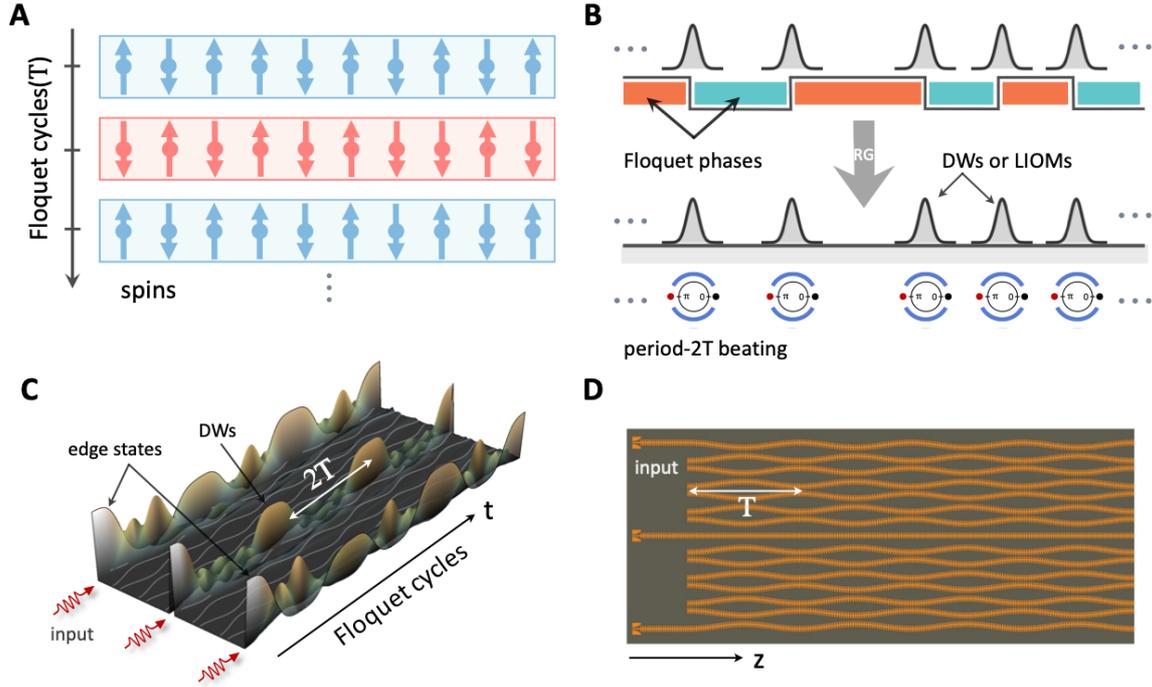

**Fig. 1. Period-$2T$ Floquet time-crystalline phases in topologically protected single-particle pictures.** (**A**) Many-body-localization-enabled Floquet time crystal on a spin chain. (**B**) Emergence of domain walls (DWs) or local integrals of motions (LIOMs) in one-dimensional disordered Floquet phases. The superposition of Floquet topological invariants zero and $\pi$ modes of the renormalized low-lying DW excitations offers the rigid period-doubling beating as a time-crystalline clock. (**C**) Schematic of the stroboscopic evolutions of DWs and edge states on a chain (e.g., spins, atoms, and waveguides) with a periodically driven protocol, demonstrating the pertinent period-$2T$ beating. (**D**) Schematic of the curved waveguide array structure with three field inputs.

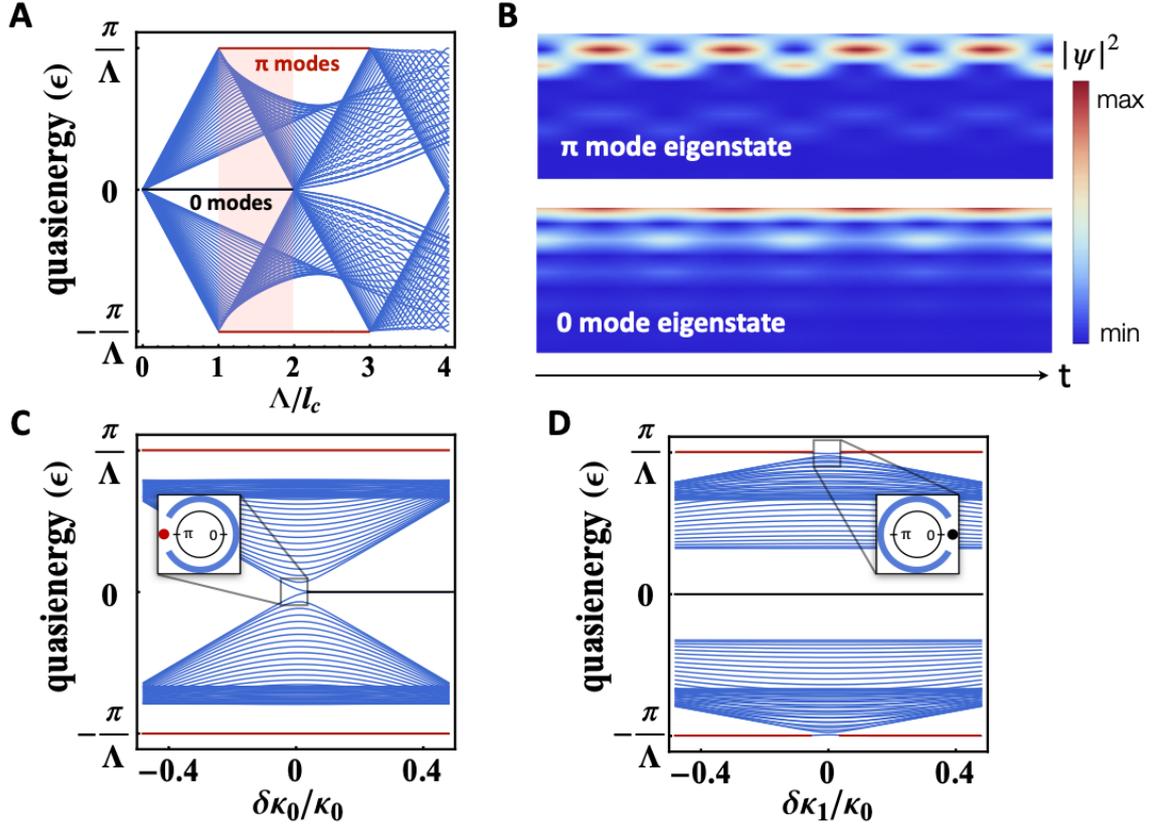

**Fig. 2. Quasienergy spectrum of photonic FTC and micromotion eigenstates of topological 0 and $\pi$ modes.** (**A**) Quasienergy band as a function of the curving period ($\Lambda$). (**B**) Micromotions of $\pi$-mode and 0-mode eigenstates in four Floquet cycles ($4T$). Floquet cycle $T$ is mapped to the curving period $\Lambda$. The two Floquet topological modes coexist in the driven condition $\Lambda/l_c \in (1, 2)$, and the coupling length $l_c = \pi/2\kappa_0$. The parameters are $\kappa_0 = 0.25$, $\delta\kappa_0 = 0.06$, and $\delta\kappa_1 = 0.12$. (**C**) Zero-gap as a function of global dimerization $\delta\kappa_0/\kappa_0$, closed at $\delta\kappa_0/\kappa_0=0$. (**D**) $\pi$-gap as a function of time-period dimerization $\delta\kappa_1/\kappa_0$, closed at $\delta\kappa_1/\kappa_0 = 0$. The insets of (C) and (D) are the quasienergy bands of the solitary existence of $\pi$ and 0 modes at $\Lambda/l_c = 4/3$, respectively.

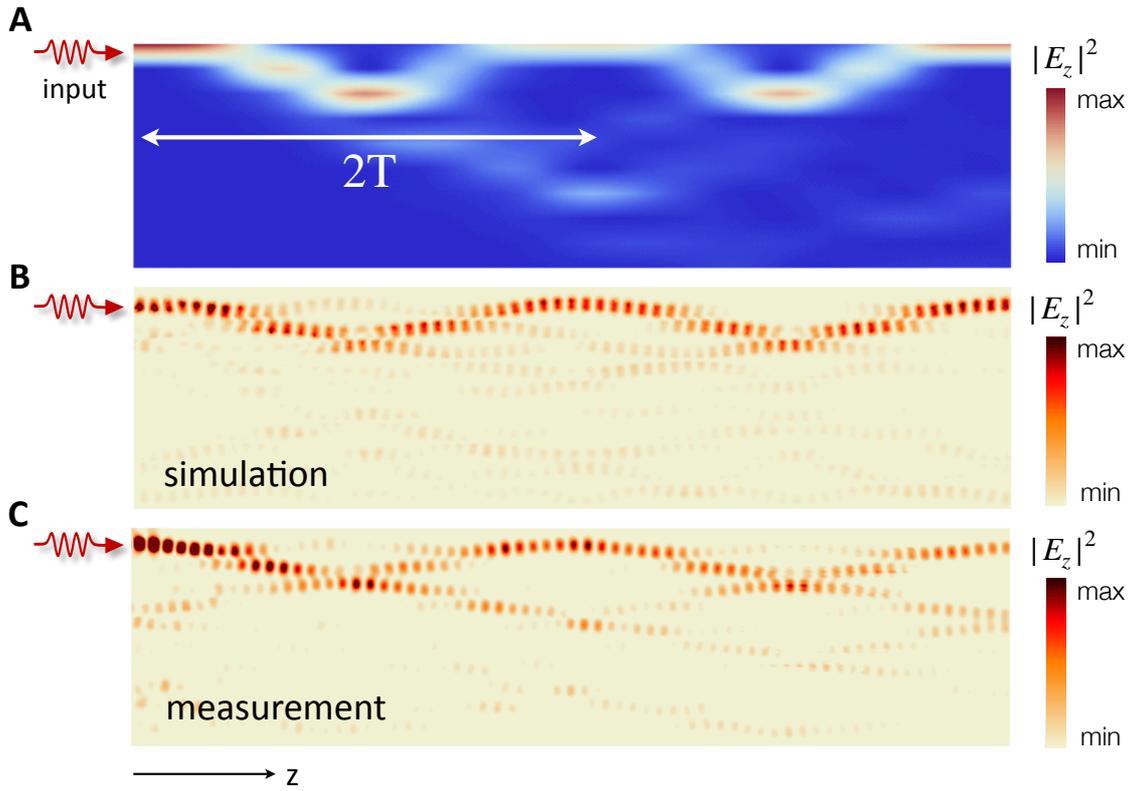

**Fig. 3. Observation of period-$2T$ oscillation in photonic Floquet simulator.** (**A**) Theoretical calculation based on the Floquet evolution operator. (**B**) FEM simulation based on COMSOL. (**C**) Near-field measurement based on a fabricated array sample. The field was input from the edge of the waveguide array (waveguide number $N = 10$) with an input frequency of 17.0 GHz for the simulations and 16.9 GHz for the experiments. The length of the simulator is $L = 400$ mm. The theory, simulation, and experiment are in perfect agreement with each other. The corresponding parameters are presented in the main text.

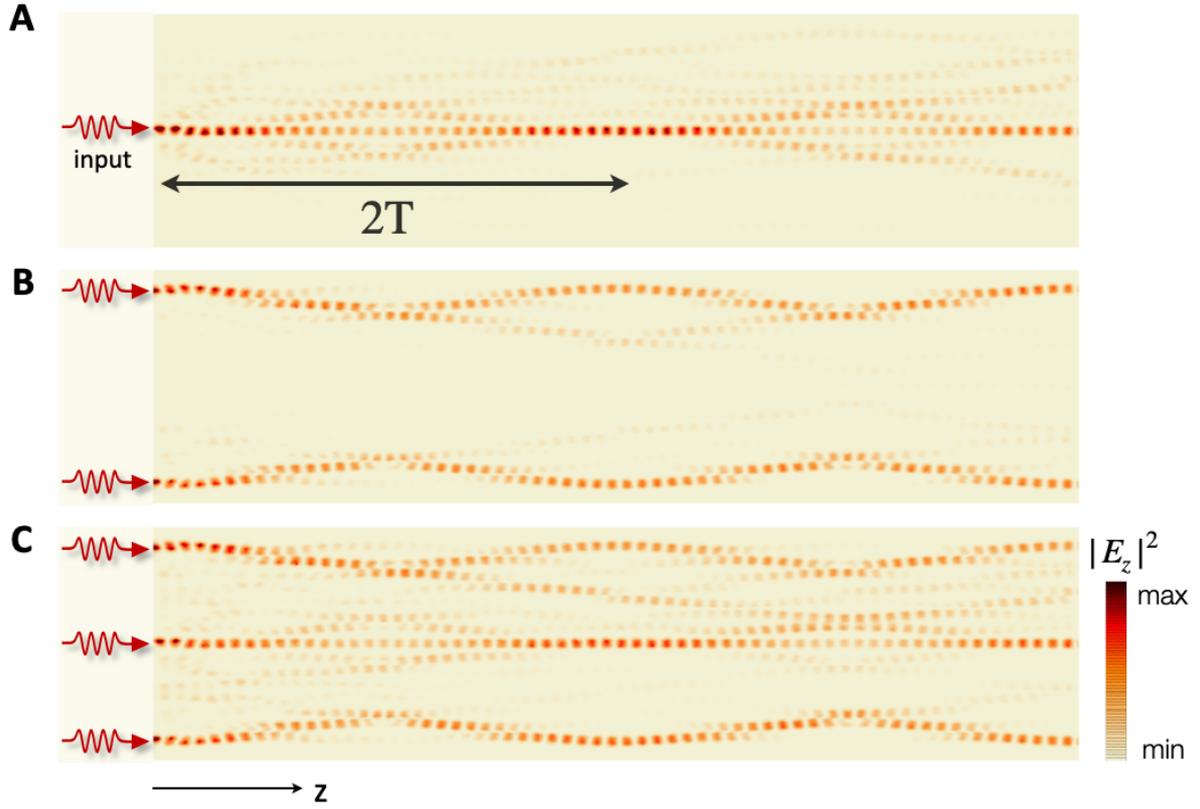

**Fig. 4. Comparison between single-particle and many-body time crystals with different inputs from edge states and a central DW.** (**A**) Near-field evolution with the central DW input. (**B**) Near-field evolution with two edge-state inputs. (**C**) Near-field evolution with combine edge-state and domain-wall inputs. Simulations based on the samples (Fig. 1D) demonstrate the photonic period-$2T$ oscillation. The structural parameters are $N = 15$, $L = 400$ mm, and $\Lambda = 100$ mm, and the input frequency is 17.0 GHz.

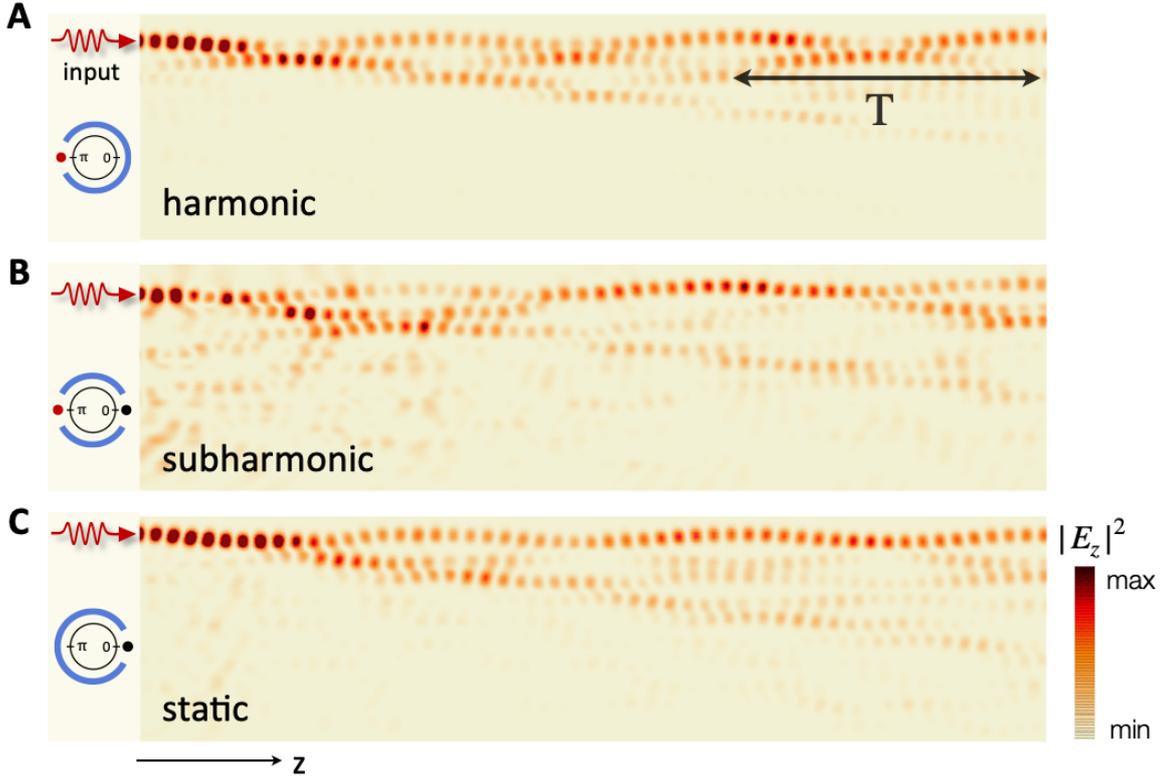

**Fig. 5. Observation of topological phase transition in Floquet simulators.** By altering the ratio between the periodic and global dimerization spacing, we can open the corresponding quasienergy gap, as shown in the insets. (**A**) For $G_{min1} = 0.9$ mm and $G_{min2} = 0.9$ mm, only $\pi$ mode exists, and the photonic system exhibits the period-$T$ oscillation when the $\pi$ mode is excited. (**B**) For $G_{min1} = 1.9$ mm, $G_{min2} = 0.9$ mm, 0 and $\pi$ modes coexist, and the system exhibits period-$2T$ oscillation when the two modes are excited simultaneously. (**C**) For $G_{min1} = 2.3$ mm and $G_{min2} = 0.9$ mm, only 0 mode is effectively excited. The system exhibits nondriven static behavior. Note that the input frequency is 17.4 GHz, array length $L = 300$ mm, and period $\Lambda = 100$ mm; therefore, the Floquet cycle is $3T$.

Supplementary Materials for

# Photonic Floquet time crystals

Bing Wang[†,1], Jiaqi Quan[†,2], Jianfei Han[3], Xiaopeng Shen[*,3], Hongwei Wu[*,2], and Yiming Pan[*,4]

*Corresponding author. Email: yiming.pan@campus.technion.ac.il; hwwu@aust.edu.cn; xpshen@cumt.edu.cn

**This PDF file includes:**

Materials and Methods

Supplementary text.

Figs. S1-S6.

# Materials and Methods

## 1. Sample fabrication, setup, and near-field measurement

The sample was fabricated on a printed circuit board (PCB, F4BK) with a dielectric constant of 2.65, loss tangent of 0.001, thickness of 0.2 mm, copper film thickness of 0.018 mm, and total array length of 400 mm. To eliminate reflections, we added an additional coplanar waveguide coupler of 50 mm at beginning of the waveguide array for guiding the input signal. To assess the array design, we fabricated several samples with different sizes and measured the evolutionary field distribution using a near-field scanner. Using the standard PCB manufacturing process, we printed several types of waveguide arrays on the same dielectric substrate with a curving period $\Lambda = 100$ mm, $G_{min1} = 0.9$ mm, and $G_{min2} = 0.9$ mm (for $\pi$ modes only), and $\Lambda = 100$ mm, $G_{min1} = 1.9$ mm, and $G_{min2} = 0.9$ mm (for both zero and $\pi$ modes). The spacing profiles are illustrated in the following section. Note that $G_{min1}$ is the minimal spacing between the first and second curved waveguides in the array, and $G_{min2}$ the minimal spacing between the second and third curved waveguides.

Near-field scanning technology was used to map the electric field around the sample surface. The instrumentation used included vector network analyzer (KeysightE5063A), translation platform, two-dimensional platform controller, detector, and coaxial transmission line (Fig. S1 of Supplementary Material). The sample was pasted on a foam that was larger than the array. The foam was placed on a mobile bottom plate platform. The bottom plate was allowed to move in two directions, $X$ and $Y$, in a step length of 0.1 mm and was controlled by the two-dimensional platform controller. One port of the vector network analyzer was connected to a feed coaxial line for microwave source signals. The other port was connected to a coaxial detection probe. The probe was fixed on a fixed frame, and the inner conductor of the monitoring probe was extended by 1 mm to observe the $Z$ component of the electric field in the plane 1 mm above the sample. A two-dimensional mapping code was used to analyze the electric near-field information, the results of which were compared with those of the simulations. The results showed that, in the near-field distribution of $E_z$ component, for sample 1 with $N = 4$, $G_{min1} = 0.9$ mm, and $G_{min2} = 0.9$ mm, the simulation and measurement results converged, and a period-$T$ $\pi$-mode propagation appeared. For sample 2 with $N = 4$, $G_{min1} = 1.9$ mm, and $G_{min2} = 0.9$ mm, a period-$2T$ oscillation behavior appeared (Fig. 3).

## 2. Numerical simulations

To verify the theoretical prediction, we conducted full-wave simulations for TM-polarized waves propagating along the $z$-direction based on finite element analysis on COMSOL Multiphysics. The spoof plasmonic waveguide arrays in the microwave regime were printed on a dielectric substrate F4BK with the following parameters: a thin spoof plasmonic waveguide made of a copper strip with a thickness of 0.3 mm, a total array length of $L = 400$ mm. The parameters of the structural unit cell ("H"-bar) were given by, $P = 2$ mm, $H = 4$ mm, $a = 0.8$ mm, and $W = 1$ mm (Fig. S1). The biatomic spoof plasmonic waveguides were designed in conformance to the relation $G(z) = g_0 \pm g_1 \pm 2A_0 \cos(2\pi z/\Lambda + \theta)$, where $g_0$ is the average spacing between waveguides and $g_1$ is the global dimerization spacing. In addition, $A_0 = 0.8$ mm is the fixed $z$-periodic dimerization spacing with the period $\Lambda$ and initial phase $\theta$, which is cosine-modulated in the propagation direction $z$. The coupling length $l_c = \pi/2\kappa_0$ and coupling strength $\kappa_0$ as a function of the spacing were simulated, as shown in Fig. S2.

In particular, to simulate the time-crystalline period–$2T$ beating with the array dimerization of spoof plasmonic waveguides, we set the minimum internal spacing of two adjacent waveguides $(G_{min1}, G_{min2}) = (0.9 \text{ mm}, 0.9 \text{ mm})$ for time-periodic dimerization only, which corresponded to $g_0 = 2.5 \text{ mm}, g_1 = 0, A_0 = 0.8 \text{ mm}$, and $(G_{min1}, G_{min2}) = (1.9 \text{ mm}, 0.9 \text{ mm})$ for both the global and periodic dimerizations which corresponded to $g_0 = 3.0 \text{ mm}, g_1 = 0.5 \text{ mm}, A_0 = 0.8 \text{ mm}$. The total number of bending periods (Floquet cycles) within $L$ for all waveguides was set to $N = 4$. The drive signal with a frequency of 17.0 GHz was input from the first waveguide in this simulation. The boundary of the whole array structure was defined as the scattering boundary condition.

## 3. Theoretical calculations

**Mapping ½-spin chain to SSH chain** - The one-dimensional ½-spin chain system studied in the $\pi$SG/DTC phase (*7*) is given by

$$H = -\sum_{i=1}^{N-1} J_i(t)\sigma_i^x \sigma_{i+1}^x - \sum_{i=1}^{N} h_i(t)\sigma_i^z + J_z \sum_{i=1}^{N-1} \sigma_i^z \sigma_{i+1}^z. \tag{M1}$$

For simplification, we averaged the disorders and time periodicity and considered the small interaction term,

$$\bar{J} = \langle\langle J_i(t)\rangle\rangle, \bar{h} = \langle\langle h_i(t)\rangle\rangle, \text{ and } J_z \ll \bar{J}, \bar{h}. \tag{M2}$$

If $\bar{J} > \bar{h}$, then the first term dominates, which leads to a $Z_2$ breaking ferromagnetic phase (FM) (i.e., spin glass, at $J_z \neq 0$). If $\bar{J} < \bar{h}$, the second term dominates, which leads to a paramagnetic phase (PM). The critical point at $\bar{J} = \bar{h}$ corresponds to the FM–PM phase transition.

We then performed the Jordan–Wigner transformation

$$\sigma_i^+ = 2\exp\left[-i\pi \sum_{k=1}^{i-1} c_k^\dagger c_k\right] c_i^\dagger, \sigma_i^- = 2\exp\left[i\pi \sum_{k=1}^{i-1} c_k^\dagger c_k\right] c_i, \tag{M3}$$

where $\sigma_i^\pm = \sigma_i^x \pm i\sigma_i^y$ are the raising and lowering operators, $c_i^\dagger$ and $c_i$ are the creation and annihilation operators of fermions at site $i$. Through transformation (M3), we mapped the spin chain to a $p$-wave pairing superconductor with the near-nearest Coulomb interaction as

$$H = -\sum_{i=1}^{N-1} J_i(t)(c_i^\dagger - c_i)(c_{i+1}^\dagger + c_{i+1}) - \sum_{i=1}^{N} h_i(t)(2c_i^\dagger c_i - 1) \\ + J_z \sum_{i=1}^{N-1}(2c_i^\dagger c_i - 1)(2c_{i+1}^\dagger c_{i+1} - 1). \quad (M4)$$

Furthermore, the Dirac fermionic creation and annihilation operators $c_i^\dagger$ and $c_i$ can be written as the Majorana operators $c_i = \frac{1}{2}(\gamma_{2i-1} - i\gamma_{2i})$ and $c_i^\dagger = \frac{1}{2}(\gamma_{2i-1} + i\gamma_{2i})$. It is easy to verify that the Majorana fermions satisfy the relations $\gamma_i = \gamma_i^\dagger$, $\{\gamma_i, \gamma_j\} = 2\delta_{i,j}$. Thus, we expressed the spin Hamiltonian (M1) in terms of the Majorana fermions:

$$H = -i\sum_{i=1}^{N-1} J_i(t)\gamma_{2i}\gamma_{2i+1} - i\sum_{i=1}^{N} h_i(t)\gamma_{2i}\gamma_{2i-1} - J_z \sum_{i=1}^{N-1}(\gamma_{2i}\gamma_{2i-1})(\gamma_{2i+2}\gamma_{2i+1}). \quad (M5)$$

where $h_i(t)$ is the intra-coupling strength at site $i$ and $J_i(t)$ is the inter-coupling strength between site $i$ and $i+1$. Thus, it is a dimerized Majorana chain with long-range interaction.

Note that, prior to the appearance of topological edge states and DWs in the dimerized chain, we suppressed the interaction terms ($J_z = 0$) and disorders. Indeed, the main objective of introducing disorders is to create DW excitations in the chain by purpose, while that of the small interaction is to add the renormalized kink–kink interaction between DWs to dephase the isolated DW product many-body state into a fully entangled many-body state. The corresponding Kitaev model for the clear, noninteracting spin model is then given by

$$\bar{H} = -i\bar{J} \sum_{i=1}^{N-1} \gamma_{2i}\gamma_{2i+1} - i\bar{h} \sum_{i=1}^{N} \gamma_{2i}\gamma_{2i-1}. \quad (M6)$$

This is an analogy of the celebrated SSH model in the Majorana sublattice basis. Majorana dimerization occurs at $|\bar{J}| \neq |\bar{h}|$. If $|\bar{J}| > |\bar{h}|$. Two unpaired Majorana zero modes appear at the ends ($i = 1$, and $2N + 1$), corresponding to the $Z_2$ breaking ferromagnetic phase in the original Ising model (M1). If $|\bar{J}| < |\bar{h}|$, the ground state will be trivially gapped, corresponding to the paramagnetic phase.

Unlike Khemani et al.'s drive protocol (*7*), the driven parameters of the setting in this study are continuously configured as

$$\begin{aligned} \bar{J}(t) &= \kappa_0 + \delta\kappa_0 + \delta\kappa_1 \cos(\omega t + \phi), \\ \bar{h}(t) &= -(\kappa_0 - \delta\kappa_0 - \delta\kappa_1 \cos(\omega t + \phi)). \end{aligned} \quad (M7)$$

We can then rewrite the time-periodic Hamiltonian (M6) as the Majorana SSH-type

$$\bar{H}(t) = -i \sum_{i=1}^{2N-1} [\kappa_0 + (-1)^i (\delta\kappa_0 + \delta\kappa_1 \cos(\omega t + \phi))] \gamma_i \gamma_{i+1}. \quad (M8)$$

We can also rewrite (M8) as the anti-symmetric skew-Hermitian matrix form $H = -\frac{i}{2} \sum_{i,j=1}^{2N-1} A_{i,j} \gamma_i \gamma_j$ and numerically solve its quasienergy spectrum. The new expression shares a similar quasienergy spectrum as the driven SSH model, as observed in the waveguide arrays.

**Floquet theory and quasienergy spectrum -** The time evolution of a time-dependent system is governed by the Schrödinger equation ($\hbar = 1$)

$$i\partial_t |\psi(t)\rangle = H(t)|\psi(t)\rangle, \quad (M9)$$

where $|\psi(t)\rangle = U(t, t_0)|\psi(t_0)\rangle$ is the state of the system at $t$, and $|\psi(t_0)\rangle$ at $t_0$. The time evolution operator $U(t, t_0)$ is the solution to the Cauchy problem $i\partial_t U(t, t_0) = H(t)U(t, t_0)$, with the initial value being $U(t_0, t_0) = 1$. The operator can be expressed as

$$U(t, t_0) = \hat{T} e^{-i \int_{t_0}^{t} H(t')dt'}$$

where $\hat{T}$ is a time-ordering operator. To numerically calculate the operator, we divided the evolution operation into an infinite series

$$U(t, t_0) = \lim_{\Delta t \to 0} e^{-iH(t-\Delta t)\Delta t} e^{-iH(t-2\Delta t)\Delta t} \ldots e^{-iH(\Delta t)\Delta t} e^{-iH(0)\Delta t}. \tag{M10}$$

We can then obtain the dynamic evolution of the time-dependent system (Fig. 3A).

For a periodically driven system with a drive period $T$, the Hamiltonian of the system has a discrete time-translation symmetry $H(t + T) = H(t)$. The Floquet theorem states that the solution of the Schrödinger equation of a periodically driven system takes the form of (*42, 43*)

$$|\psi_n(t)\rangle = e^{-i\epsilon_n t}|u(t)\rangle, \quad |u_n(t)\rangle = |u_n(t + T)\rangle, \tag{M11}$$

where $|\psi_n(t)\rangle$ is the Floquet state, $|u_n(t)\rangle = |u_n(t + T)\rangle$ is the periodic Floquet mode, and $\epsilon_n$ is the quasienergy. Considering the time periodicity, the time evolution operator of the system can be written as $U(t, t_0) = U(t, t_0 + nT)[U(t_0 + T, t_0)]^n$, where $U(t_0 + T, t_0)$ is the Floquet operator defined over one driving period.

Through the Floquet operator, we defined an effective time independent Floquet Hamiltonian $U(t_0 + T, t_0) = e^{-iH_{F[t_0]}T}$, where $t_0$ is the Floquet gauge. By diagonalizing the effective Hamiltonian with the driven tight-binding Hamiltonian, we obtain

$$H_{F[t_0]} = \frac{i}{T} \log U(t_0 + T, t_0). \tag{M12}$$

We calculated the quasienergy spectrum and the Floquet eigenstates as in Fig. 2 of the main text, in which the gauge $t_0 = 0$ and the quasienergy ($\epsilon$) modulo $2\pi/T$ are confined in the Floquet–Brillouin zone.

**Zero- and $\pi$-gap invariants -** To fully characterize the topological properties of the Floquet system by the gap invariants (*41*), we considered the micromotion operator

$$V(t, t_0) = U(t, t_0) e^{i H_{F[t_0]}(t-t_0)}, \tag{M13}$$

which describes the dynamic evolution within each driving cycle. Under the periodic boundary condition, the Hamiltonian of the driven SSH model in momentum space can be written as

$$H(k,t) = \left(\kappa_0 + \delta\kappa_0 + \delta\kappa(t) + \left(\kappa_0 - \delta\kappa_0 - \delta\kappa(t)\right)\cos(k)\right)\sigma_x$$
$$+ \left(\kappa_0 - \delta\kappa_0 - \delta\kappa(t)\right)\sin(k)\,\sigma_y, \tag{M14}$$

where $\delta\kappa(t) = \delta\kappa_1 \cos(\omega t)$ is the time-periodic staggered coupling and $T = 2\pi/\omega$ is the driving period. The Hamiltonian has a chiral symmetry, which is defined as the unitary chiral operator $\Gamma = \sigma_z$,

$$\Gamma H(t,k)\Gamma^{-1} = -H(-t,k), \tag{M15}$$

and the chiral symmetry has a constraint on the micromotion operator $\Gamma V_{\epsilon T}(t,k)\Gamma^{-1} = -V_{-\epsilon T}(-t,k)e^{i2\pi t/T}$. For zero modes, $\epsilon = 0$ and $t = T/2$, we noted that $\Gamma V_0(T/2,k)\Gamma^{-1} = -V_0(T/2,k)$, which is anti-diagonal on the chiral basis,

$$V_0(T/2,k) = \begin{pmatrix} 0 & V_0^+ \\ V_0^- & 0 \end{pmatrix}. \tag{M16}$$

The chiral invariant is defined by

$$G_0 = \frac{i}{2\pi}\int_{-\pi}^{\pi} tr((V_0^+)^{-1}\partial_k V_0^+)dk. \tag{M17}$$

For $\pi$ mode, $\epsilon = \pi/T$ and $t = T/2$, we obtained $\Gamma V_\pi(T/2, k)\Gamma^{-1} = V_\pi(T/2, k)$, which is a diagonal in the chiral basis,

$$V_\pi(T/2, k) = \begin{pmatrix} V_\pi^+ & 0 \\ 0 & V_\pi^- \end{pmatrix}. \tag{M18}$$

The chiral invariant is defined by

$$G_\pi = \frac{i}{2\pi} \int_{-\pi}^{\pi} tr((V_\pi^+)^{-1} \partial k\, V_\pi^+) dk. \tag{M19}$$

As a result, we found that for the driven SSH Hamiltonian (M14), the invariant $G_{0(\pi)} = 1$ corresponded to the Floquet eigenstates at the quasienergy $0(\pi)$-gap, while $G_{0(\pi)} = 0$ corresponded to no such eigenstate.

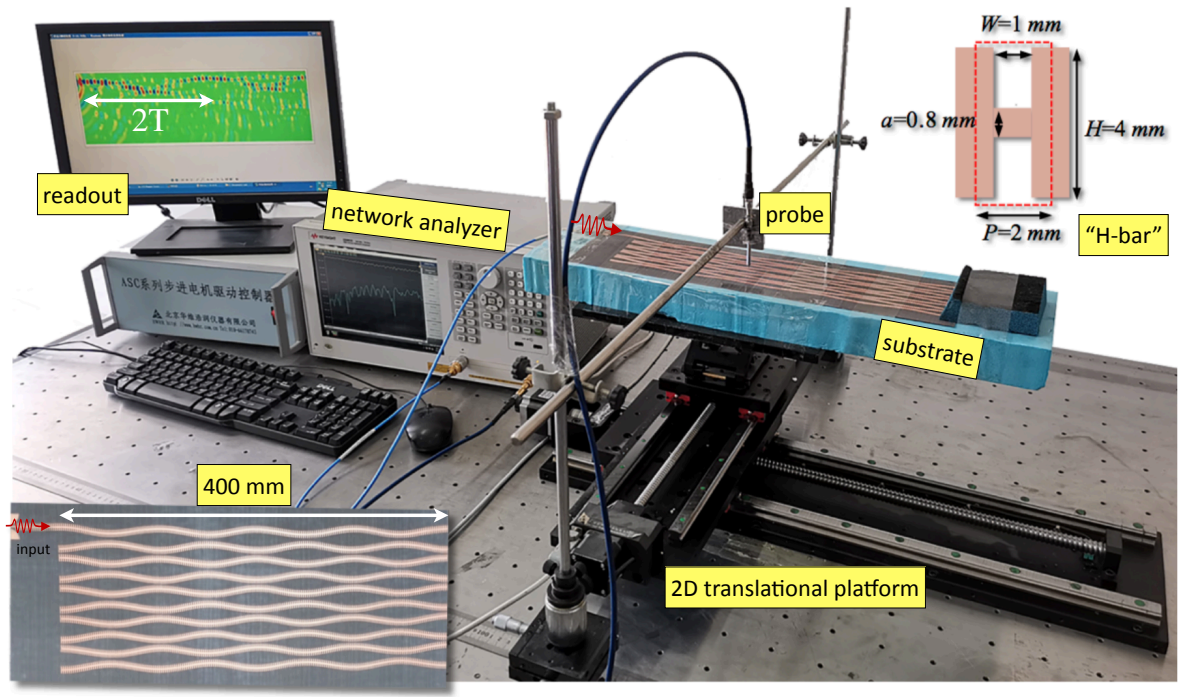

**Fig. S1. Photograph of the microwave near-field experiment platform**. The total experiment platform includes vector network analyzer (KeysightE5063A), two-dimensional translation platform, and platform controller. The array sample is pasted on a foam whose dielectric constant near 1 has hardly influence on the sample. The foam is placed on the movable bottom plate platform, which can move in the *X*-, and *Y*-directions controlled by a mechanical stage. One port of the vector network analyzer is connected to the feed coaxial line to provide the microwave source signal, whereas the other port is connected to the coaxial detection probe (the tip). The probe is fixed on the fixed frame to map the *Z* component of the electric field in the plane 1 mm above the sample. Field density information can be observed by a specific computer program. The lower left inset is an enlarged photograph of the sample.

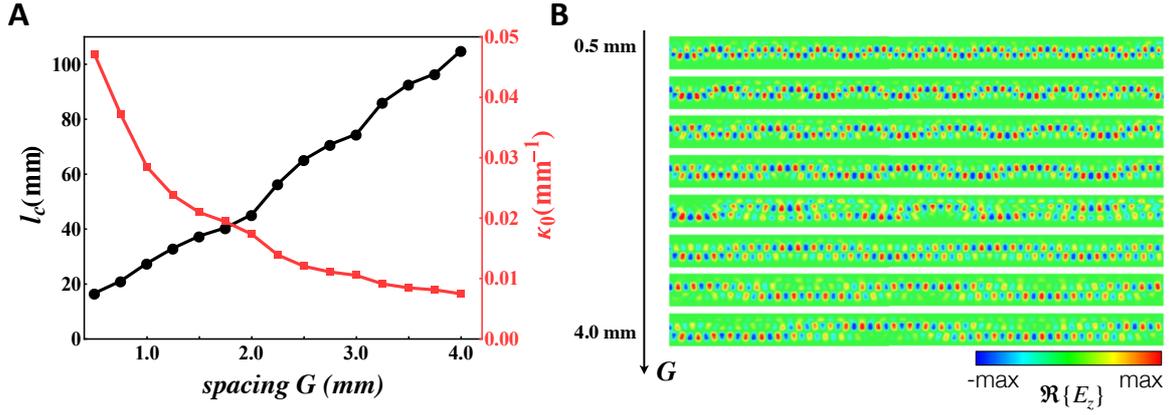

**Fig. S2. Reconstruction of coupling profiles from the simulation of two coupled waveguides.** (**A**) Coupling length $l_c$ and the coupling strength $\kappa_0$ as functions of the spacing ($G$) of the adjacent waveguides. The relation between coupling length and coupling strength is given by $l_c = \pi/2\kappa_0$. (**B**) Simulation results of the electric-field propagation between two coupled waveguides. The spacing of the adjacent waveguides varies from 0.5 mm to 4.0 mm. The coupling length can be estimated by the effective distance of the input field fully propagating from one waveguide to the next. Note that the waveguide length is 400 mm, and the input frequency used for simulation is 17.0 GHz.

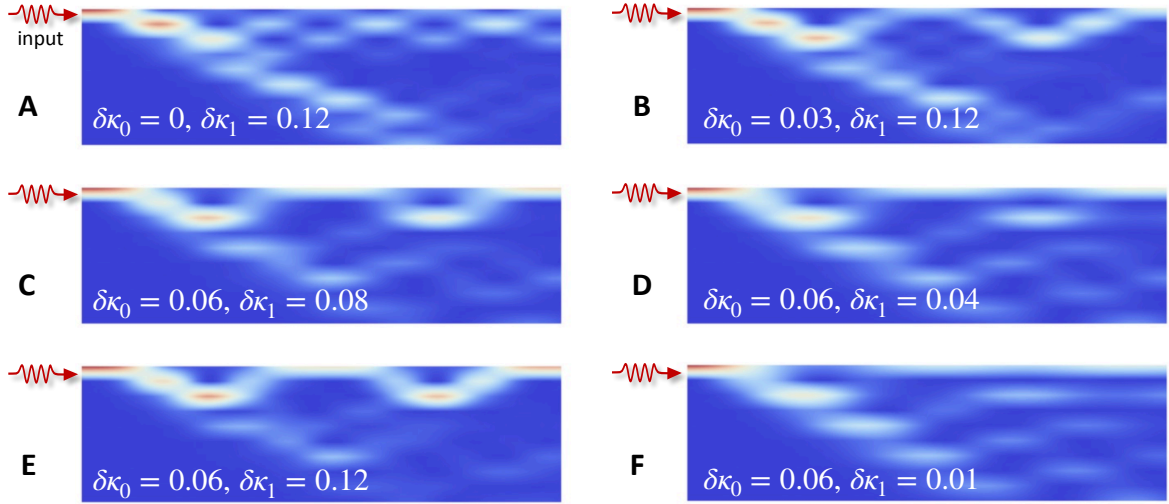

**Fig. S3. Theoretical results for the Floquet topological phase transition with varying dimerized coupling strength.** (A) The only $\pi$ mode phase with $\delta\kappa_0 = 0, \delta\kappa_1 \neq 0$. (**B-E**) Time-crystalline phase coexistence with $\delta\kappa_0 \neq 0, \delta\kappa_1 \neq 0$. (**F**) Zero mode phase with $\delta\kappa_0 \neq 0, \delta\kappa_1 = 0$. The theoretical results are calculated with 80 waveguides and $\omega/\Delta = 0.75, \kappa_0/\Delta = 0.25, \Delta = 1$ (bandwidth). Note that for high-resolution demonstration, we plot only the first ten waveguides of the array in four Floquet cycles (4$T$).

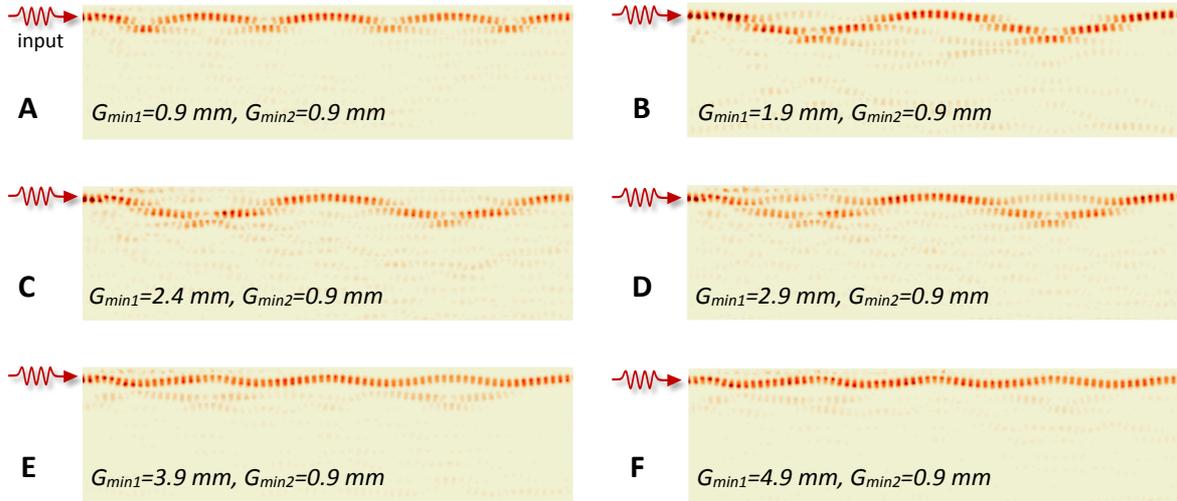

**Fig. S4. Simulation results for the Floquet topological phase transition with changing adjacent waveguide spacing.** (**A**) $\pi$-mode phase. (**B-D**) Time-crystalline phase. (**E, F**) Zero-mode phase. There are ten waveguides in an array, and the total length of the array is $L = 400\ mm$, with a curving period of $100\ mm$. The input frequency used for simulation is 17.0 GHz.

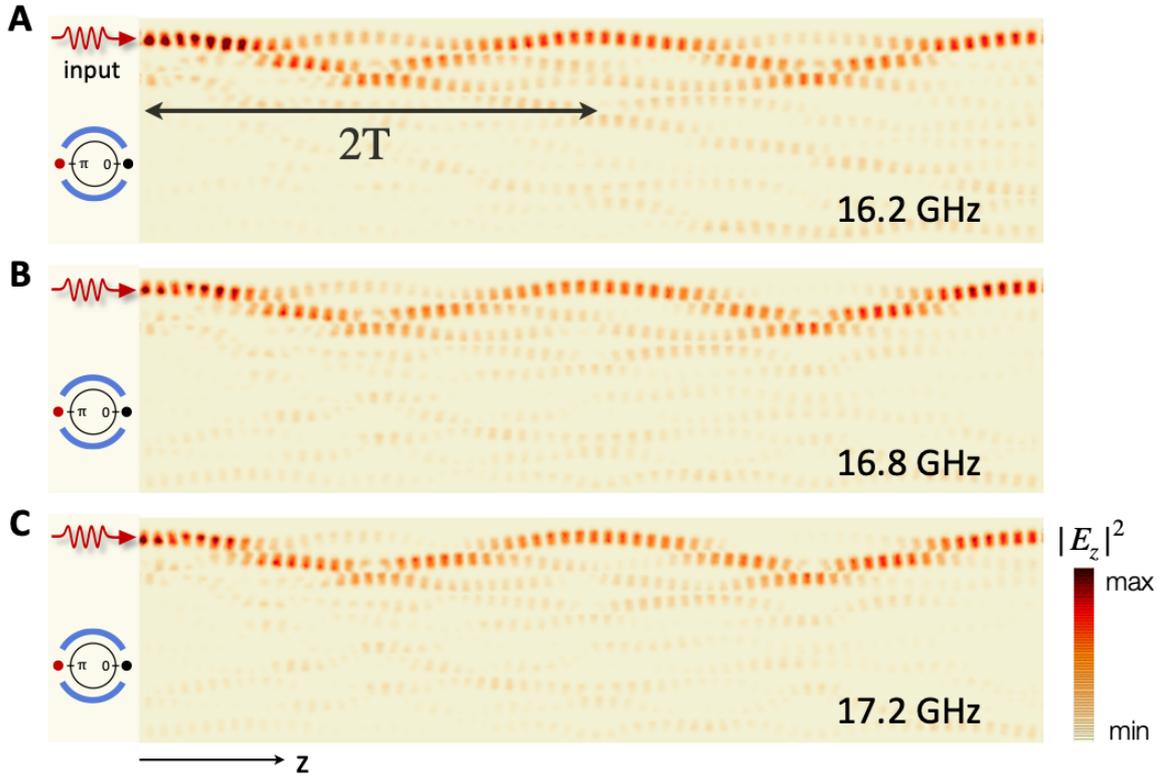

**Fig. S5. Simulation results for period-2T oscillation with different input frequencies.** (**A**) 16.2 GHz. (**B**) 16.8 GHz. (**C**) 17.2 GHz. There are ten waveguides in an array, and the length of the array is $L = 400$ mm with a curving period of 100 mm. The corresponding minimal spacings of the adjacent waveguides are $G_{min1} = 1.9$ mm and $G_{min2} = 0.9$ mm.

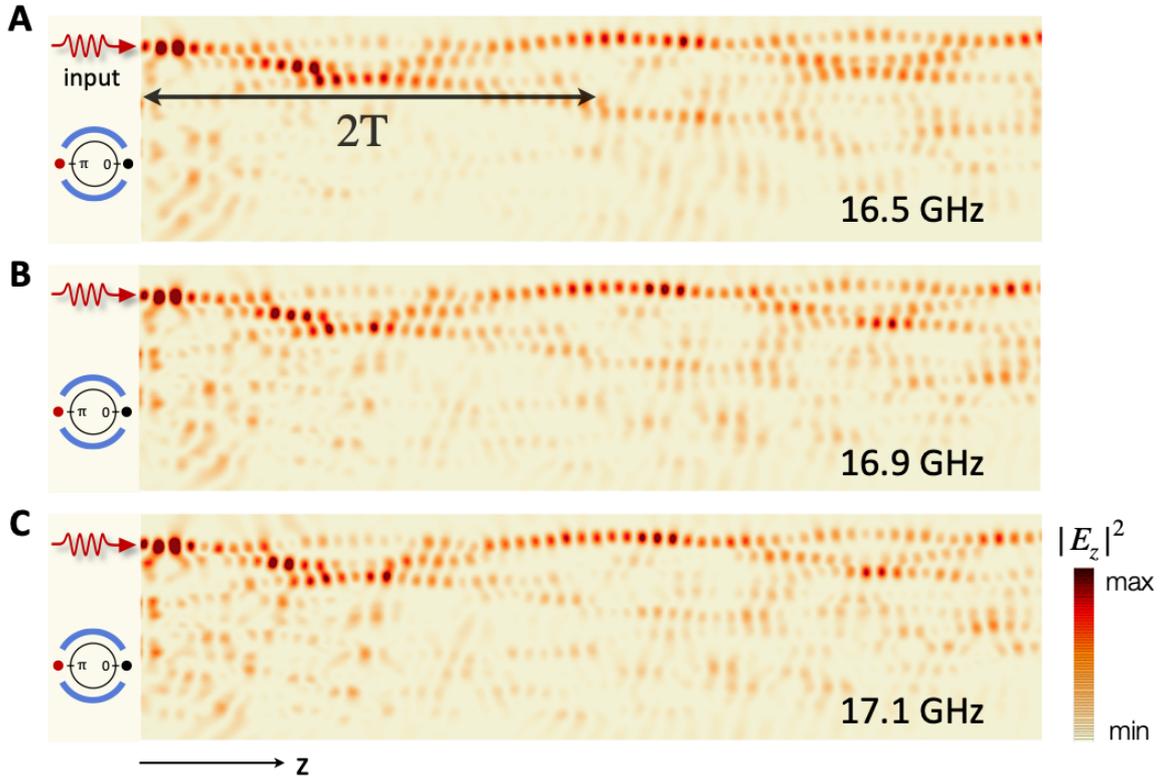

**Fig. S6. Measurement results for period-2$T$ oscillation with different input frequencies.** (**A**) 16.5 GHz. (**B**) 16.9 GHz. (**C**) 17.1 GHz. There are 10 waveguides with a total length of 400 mm and period length of 100 mm each. The minimal spacings of adjacent waveguides are $G_{min1} = 1.9$ mm and $G_{min2} = 0.9$ mm, which is the same as in Fig. S5.